\documentclass[twocolumn,preprintnumbers]{revtex4}
\usepackage{amssymb}
\usepackage{lineno}
\usepackage{graphicx}
\usepackage{soul}
\usepackage[usenames]{color}

\begin{document}

\title{An explanation of trans-ionospheric pulse pairs}

\author{H.-C. Wu}

\thanks{huichunwu@zju.edu.cn, http://mypage.zju.edu.cn/en/hcwu}

\affiliation{Institute for Fusion Theory and Simulation and Department of Physics,
Zhejiang University, Hangzhou 310027, China}

\affiliation{IFSA Collaborative Innovation Center, Shanghai Jiao Tong University,
Shanghai 200240, China}

\date{\today}

\begin{abstract}
Trans-ionospheric
pulse pairs are the most powerful natural radio signals on the Earth and associated with lightning. They have been discovered
for two decades by satellites, but their origin still remains
elusive. Here we attribute these radio signals to relativistic electrons
produced by cloud-to-ground lightning. When these electrons strike
the ground, radio bursts are emitted towards space in
a narrow cone. This model naturally explains the interval, duration,
polarization, coherence and bimodal feature of the pulse pairs. Based on electron
parameters inferred from x-ray observation of lightning, the calculated signal
intensity agrees with the measurement of satellites. Our results are useful to develop
global warning system of storms and hurricane based on GPS satellites.
\end{abstract}
\maketitle

An abrupt increase in flash rate occurs 5-20 minutes preceding
severe storms \cite{Williams} or during hurricane intensification
\cite{Shao1}. Therefore, detection of lightning as an early warning
can reduce deaths, injuries and damages related to storms or hurricanes
\cite{Hamlin}. It has been known that lightning produces electromagnetic
radiation from radio waves to gamma rays \cite{Rakov,Dwyer1}. Investigation
of lightning radiation has a fundamental interest, and also can develop novel schemes for monitoring lightning
and uncover potential radiation hazards to humans and air vehicles.

Radio bursts of 25-200MHz related to lightning have been detected
by several satellites, first the ALEXIS \cite{Holden,Massey1,Massey2,Zuelsdorf},
then the FORTE \cite{Jacobson1,Jacobson2,Light1,Tierney,Shao,Light2,Jacobson7,Jacobson3,Jacobson4,Jacobson5,Jacobson6},
and recently the Chibis-M \cite{Dolgonosov}. These signals consist of two pulses
separated by tens of $\mu$s typically and travel through the ionosphere to reach the satellites, hence they are dubbed
trans-ionospheric pulse pairs (TIPPs). These pulse pairs are much stronger than ordinary radio signals from lightning,
and occasionally exceed 1MW in power. They sometimes occur
tens of times in a storm, and the recorded events have been in excess
of half a million.

The reflection hypothesis \cite{Massey1} had been proposed to interpret
these pulse pairs. An in-cloud source is assumed to emit a radio burst with a wide pattern. The first pulse in a TIPP comes from
the direct path from the source to satellites, and the second one
corresponds to the signal reflected by the ground. The pulse interval
is determined by the source altitude ($\sim10$km) and the observation
angle. This in-cloud source \cite{Smith1} is then related to
the so-called compact intracloud discharge \cite{LeVine,Willett,Rison,Thomas,Nag}.

Although the reflection hypothesis can give the typical interval
of TIPPs, it is confronted with severe challenges. First, the hypothesis
implies that the second pulse could be weaker than the first one
due to the reflection loss. However, the mean energy ratio of two
pulses is about one \cite{Holden,Massey1,Massey2}. This requires
a surface reflectivity near 100\% \cite{Massey1}, which is not established
yet. Unexpectedly, in the sub-100MHz band, the second pulse occasionally
is a few times \cite{Massey1} or even one order of magnitude
\cite{Tierney} stronger than the first one. Moreover, in the $>110$
MHz band, the second pulse has the larger energy for more than half of
events \cite{Jacobson5}. Second, since two pulses are supposed to
originate from the same source, internal
structures of them should be congruent. However, two pulses ($>110$
MHz) are shown to be completely uncorrelated in all the events \cite{Jacobson5}.
Finally, 80\% of events occur without compact intracloud discharges
\cite{Jacobson6}.

In this paper, we propose that the TIPPs are caused by relativistic
electrons from cloud-to-ground lightning. These electrons strike the ground and induce radio emission, which is beamed towards space and
trigger the satellites. This model can explain many properties of
TIPPs. We remark that there have been evidences of associations
between the pulse pairs and cloud-to-ground lightning. In a correlation
analysis \cite{Jacobson2}, 4083 events are found correlated
with negative cloud-to-ground strokes, and only 665 events with intracloud
discharges. Moreover, the power of some pairs is shown to be proportional
to the peak current of cloud-to-ground lightning \cite{Light2}.

Relativistic electron generation \cite{Dwyer1,Dwyer2} is responsible
for x-ray emission
observed on the ground \cite{Moore,Dwyer3,Howard1,Howard2,Schaal,Mallick,Babich} from stepped leaders of cloud-to-ground
lightning. Each leader step emits an x-ray burst during the step formation. This x-ray burst can be a single spike of $\ll1\mu$s
\cite{Dwyer3} or consists of several spikes with a total duration
of $\sim1\mu$s \cite{Howard2}, corresponding to a single or multiple air breakdown for the step formation.

The consensus on the generation of relativistic
electrons is that thermal electrons are initially accelerated to keV by
an extremely intense field localized on the leader tip, and
then undergo avalanche and further acceleration in the ambient electric
fields between the leader and ground. This mechanism predicts that
the electron energy $k_{e}$ follows the Boltzmann distribution
$k_{0}^{-1}\exp(-k_{e}/k_{0})$, where $k_{0}=7.3$MeV is the average energy.
Recent analysis \cite{Babich} shows that $1\times10^{10}-4\times10^{11}$
collimating electrons can explain observed x-rays, and should be Boltzmann-distributed
at 7MeV or monoenergetic from 1 to 10MeV.

\begin{figure}[t]
\includegraphics[width=.45\textwidth]{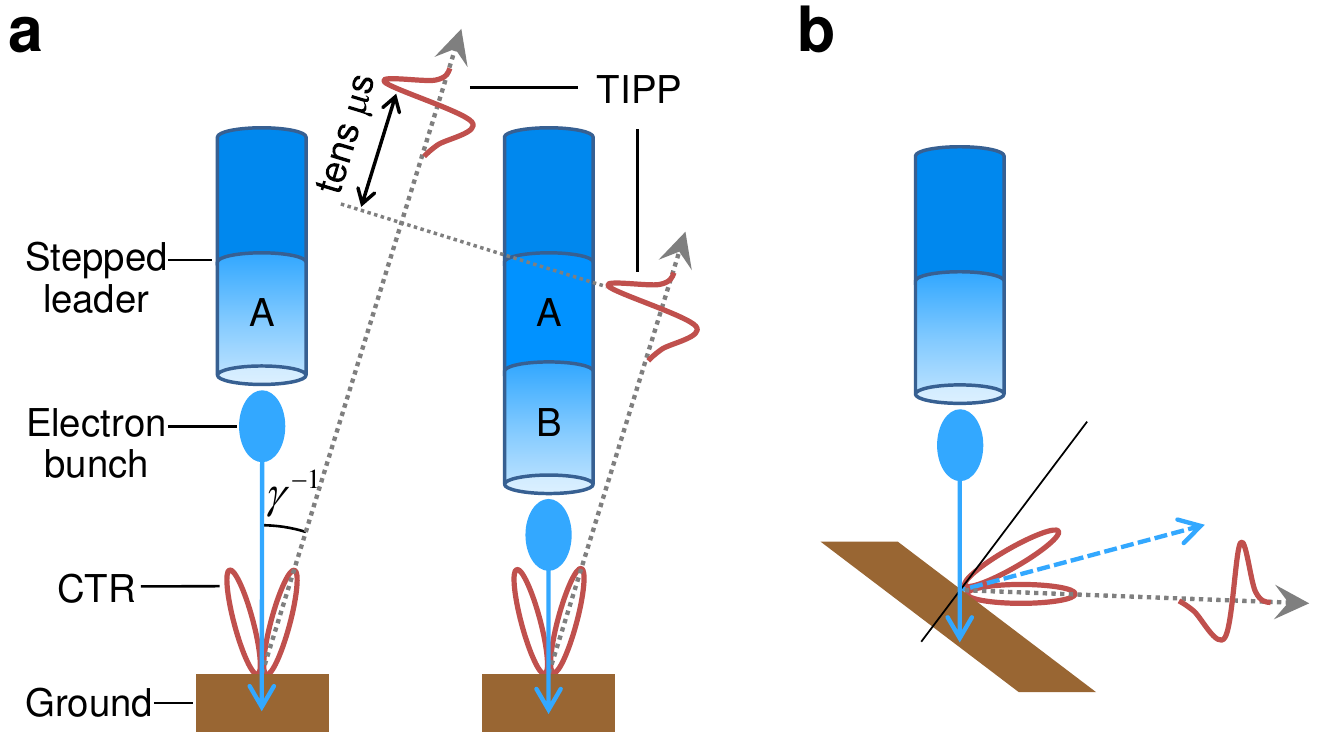} \label{fig1}
\caption{Model of TIPPs. a, Generation of a TIPP. Two groups of relativistic electrons are produced from the
last two steps A and B of a stepped leader. They strike the ground
and separately excite the first and
second pulses by coherent transition radiation (CTR). b, Scenario for the pair
detection on the ground. When electrons are obliquely incident on
a slope, the radiation can travel along the ground surface.}
\end{figure}

Relativistic electrons are expected to have the same profile as x-ray
bursts, and are organized in an isolated electron bunch of $\ll1\mu$s or a train of
electron bunches of $\sim1\mu$s totally. They are directed towards
the ground. The range of electrons (1-10MeV) in air is given by $4.4k_{eM}$,
where $k_{eM}$ is the energy in MeV. So electrons at 7MeV can
travel 31 meters in the air. The step length of the leaders varies
from 3 to 200m \cite{Rakov}. Considering the continuous acceleration
by the ambient field, relativistic electrons from the last few steps
are able to reach the ground and excite coherent transition
radiation \cite{Happek}. As sketched in Fig. 1a, two groups of electrons from successive
steps A and B account for two pulses in a TIPP. Here step B refers
to the last step nearest the ground.

To discuss the coherent transition radiation, the electron bunch is assumed
to have a Gaussian profile $n_{b0}\exp(-r^{2}/2\sigma_{t}^{2})\exp(-z^{2}/2\sigma_{l}^{2})$,
where $n_{b0}$ is the peak density, $\sigma_{t}$ and $\sigma_{l}$
are the characteristic radii in transverse and longitudinal directions,
respectively. Simulation \cite{Wu} shows
that the radiation is a bipolar pulse, and linearly-polarized
at a far-field point.

Here we utilize the new model to explain the characteristics of TIPPs.
First, the pulse separation of 7.5-$110\mu$s \cite{Holden,Massey1}
in pairs should equal the step interval, which ranges from 5 to
$100\mu$s \cite{Rakov}. Second, two pulses in pairs are either single-spike of 100ns \cite{Jacobson3} or multi-spike of totally
2-$4\mu$s \cite{Massey2}. This bimodal feature corresponds to an
isolated electron bunch ($\ll1\mu$s) or a train of bunches ($\sim1\mu$s),
as directly inferred from x-ray observation. The 100ns pulses are
linearly polarized and completely coherent, which agrees
with the coherent transition radiation for a single bunch. The 2-$4\mu$s
modulated pulses are also polarized \cite{Shao}, but incoherent. Electron
bunches in the train of $\sim1\mu$s are randomly distributed, so
no phase correlation exists among the discrete radiations of the different
bunches. During the dispersive propagation in the ionosphere, the
stretched radiations overlap and randomly interfere with each other, which destroys the coherence. Third, more electrons are expected
to reach the ground from the last step, which explains the trend
that the second pulse is more energetic. The second pulse in the 2-$4\mu$s
modulated case is more dominated by discrete and
narrow spikes \cite{Jacobson5}, which can be due to reduced travelling
diffusion of the train of electron bunches from the last step. Finally,
isolated-pulse events \cite{Massey1,Dolgonosov} can happen when only
electrons from the last step reach the ground.

\begin{figure}[t]
\includegraphics[width=.47\textwidth]{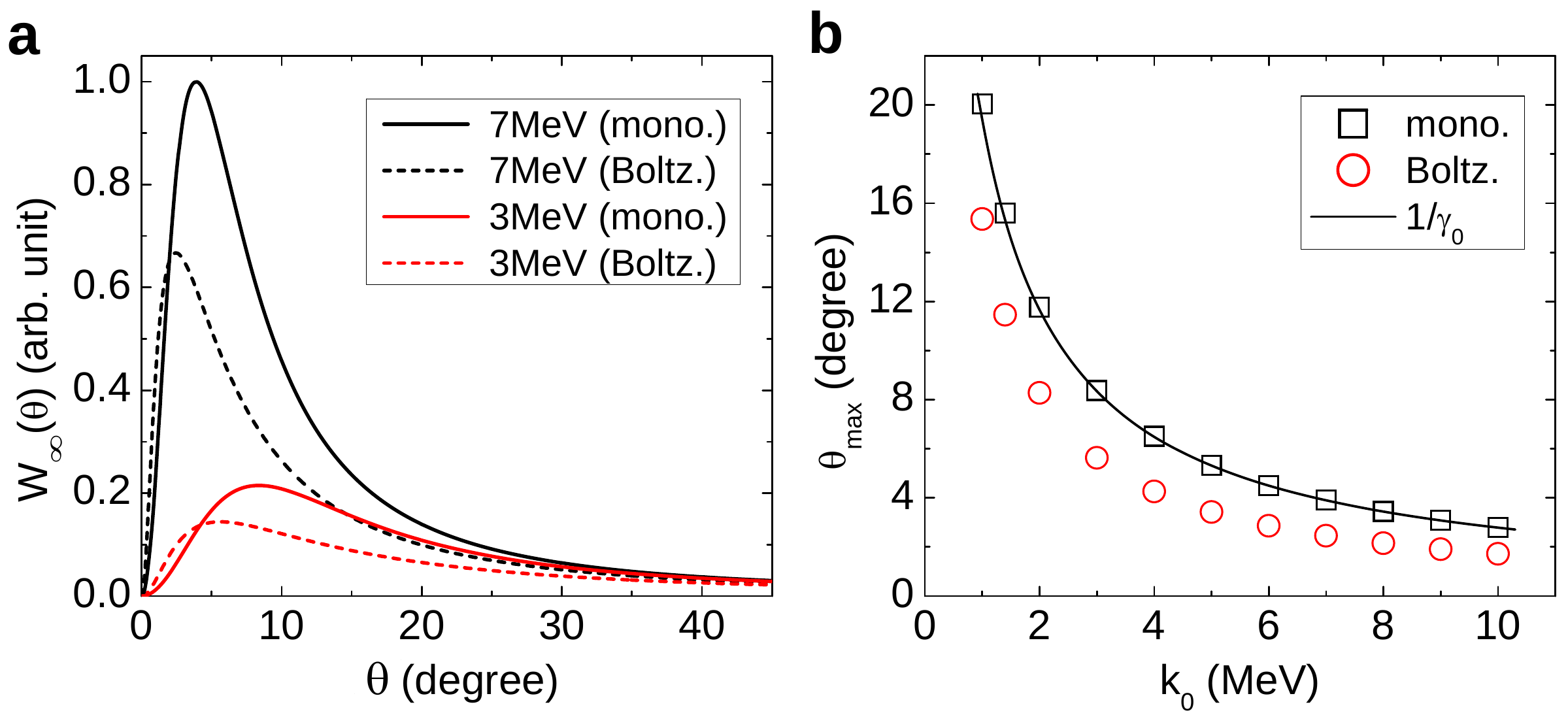} \label{fig2}
\caption{Radiation distribution. a, Angular distribution of
radiation energy for monoenergetic or Boltzmann-distributed electron
bunches with electron energies $k_{0}=3$ and 7 MeV. b, The angle of emission peak $\theta{}_{\max}$
versus $k_{0}$.}
\end{figure}

\begin{figure*}[t]
\includegraphics[width=.85\textwidth]{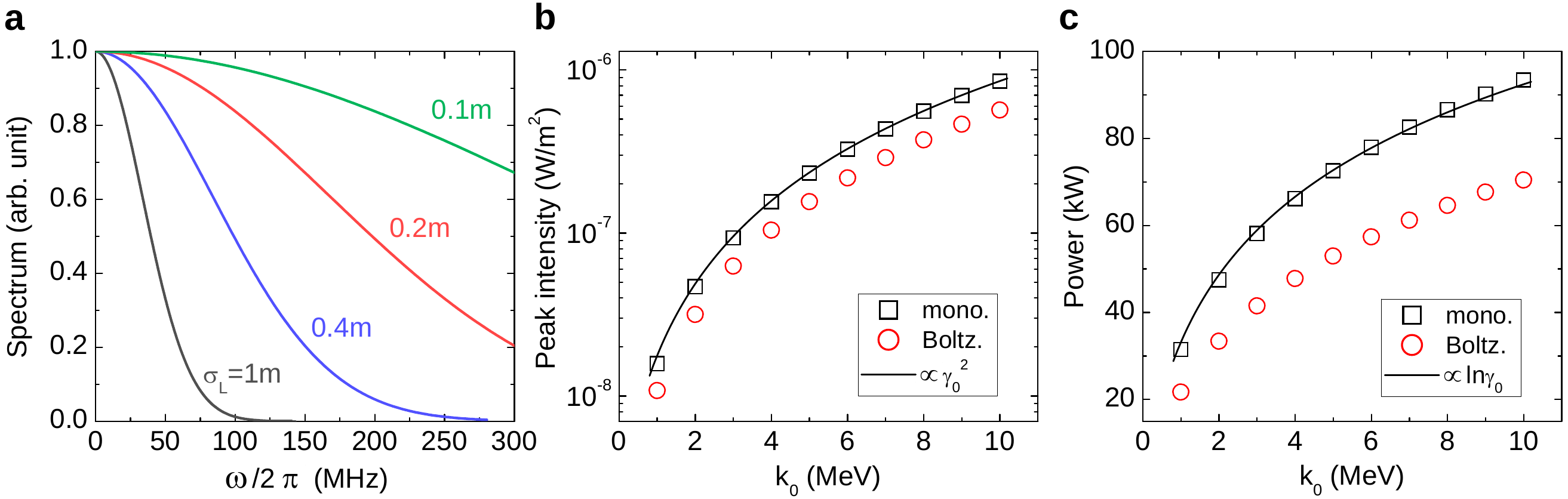} \label{fig3}
\caption{Spectrum, intensity and power of radiation. a, Spectrum
at the emission peak $\theta{}_{\max}$ for electron energy $k_{0}=7$ MeV. b, Radiation intensity at
$\theta{}_{\max}$ versus $k_{0}$. c. Radiation power versus $k_{0}$.}
\end{figure*}

Now, we quantitatively discuss the radiation pattern, spectrum and power of the TIPPs.
For a monoenergetic electron bunch normally striking the surface of
a perfect conductor with the permittivity $\varepsilon=\infty$, the
energy distribution of coherent transition radiation
in frequency and angle \cite{ZhengJ} is given by
\begin{equation}
W_{\infty}(\omega,\theta)=\frac{r_{e}mc}{\pi^{2}}\frac{N^{2}\beta_{0}^{2}\sin^{2}\theta}{(1-\beta_{0}^{2}\cos^{2}\theta)^{2}}e^{-\frac{\omega^{2}}{c^{2}}(\sigma_{t}^{2}\sin^{2}\theta+\sigma_{l}^{2}\beta_{0}^{-2})},
\end{equation}
where $\omega$ is the angular frequency, $\theta$ is the observation
angle with respect to the surface normal, $r_{e}$ and $m$ are the
classic radius and rest mass of electrons respectively, $c$ is the light speed, $N=(2\pi)^{3/2}n_{b0}\sigma_{l}\sigma_{t}^{2}$
is the total electron number, and $\beta_{0}=V_{0}/c$ is the normalized
electron velocity. For a general medium,
we have $W_{\varepsilon}(\omega,\theta)\approx\mathcal{R}(\varepsilon)W_{\infty}(\omega,\theta)$,
where $\mathcal{R}=\left\vert (\sqrt{\varepsilon}-1)/(\sqrt{\varepsilon}+1)\right\vert ^{2}$
is the Fresnel reflectivity (see Methods). In the frequency region of TIPPs,
the soil permittivity increases with its moisture $m_{s}$ \cite{Saarenketo},
and has $\varepsilon_{m_{s}=15\%}\approx10-5i$ and $\mathcal{R}\approx30\%$.
Rainfall can dramatically increase the soil moisture and its reflectivity.
There is $\mathcal{R}\approx70\%$ for sea water \cite{Meissner}.

Integrating Eq. (1) over $\omega$, we obtain the angular distribution
of radiation energy
\begin{equation}
W_{\infty}(\theta)=\frac{r_{e}mc^{2}}{2\pi^{3/2}}\frac{N^{2}\beta_{0}^{2}\sin^{2}\theta}{(1-\beta_{0}^{2}\cos^{2}\theta)^{2}}(\sigma_{t}^{2}\sin^{2}\theta+\sigma_{l}^{2}\beta_{0}^{-2})^{-\frac{1}{2}}.
\end{equation}
Figure 2a shows the radiation distribution $W_{\infty}(\theta)$ for
electron bunches with $\sigma_{t}/\sigma_{l}=1$. Radiation is
null along $\theta=0$ and maximum at the angle of $\theta_{\max}\approx1/\gamma_{0}$
(see Fig. 2b), where $\gamma_{0}=(1-\beta_{0}^{2})^{-1/2}$ is the
relativistic factor. There is $\theta_{\max}=3.9{}^{o}$ for $k_{0}=7$MeV
($\gamma_{0}\simeq14.7$), so the radiation is confined to a
very narrow cone and beamed towards space. The ratio $\sigma_{t}/\sigma_{l}$
affects the emission pattern by the term $\sigma_{t}^{2}\sin^{2}\theta+\sigma_{l}^{2}/\beta_{0}^{2}$
in Eq. (2), which is approximately proportional to $(\sigma_{t}/\sigma_{l})^{2}\gamma_{0}^{-2}+\beta_{0}^{-2}\approx1$ for $\sigma_{t}/\sigma_{l}\ll\gamma_{0}$
nearby the emission peak ($\theta=\theta_{\max}$).
At 7MeV, the radiation pattern does not change much
for $\sigma_{t}/\sigma_{l}\in[0,3]$. We also present the results
of Boltzmann-distributed bunches with the average energy $k_{0}$
(see Methods), which have a smaller $\theta_{\max}$ compared to the monoenergetic case.

The satellites record the spectrum and radiation intensity of TIPPs.
Figure 3a displays the spectrum in Eq. (1) at the emission peak
for monoenergetic bunches with $k_{0}=7$MeV and $\sigma_{t}/\sigma_{l}=1$.
The spectral profile keeps the almost same for $\sigma_{t}/\sigma_{l}\in[0,5]$
and is also not sensitive to the electron distribution function. The
spectral range is proportional to $\sigma_{l}^{-1}$, and is
broader for a shorter bunch. The TIPP spectrum can be flat within
28-166MHz \cite{Massey2}, which implies $\sigma_{l}\leq0.2$m in
Fig. 3a. Therefore, the bunch length has $L\approx4\sigma_{l}\leq0.8$m, i.e.
no longer than 3ns temporally. As discussed above, the single-spike
x-ray bursts from the stepped leader are much shorter than $1\mu$s.
It has been shown that meter-scale laboratory sparks in air \cite{Dwyer4}
emit very similar x-rays as in lightning, which are generally sub-10ns
\cite{Nguyen,Dwyer5} and can be only 1ns \cite{Kochkin}. These x-ray
observations support the generation of nanosecond electron bunches
from the stepped leader.

The radiation intensity can be calculated by $W(\theta)/(TH{}^{2}),$ where
$T\approx7.5\sigma_{l}/c$ is the radiation duration \cite{Wu} and
$H$ is the satellite altitude. Figure 3b displays the radiation intensity
at the emission peak versus $k_{0}$ for $\sigma_{l}=0.2$m, $N=5\times10^{11}$,
and $H=800$km \cite{Holden,Jacobson1}. The monoenergetic 7MeV bunch
has the peak intensity of $4.3\times10^{7}$W/m$^{2}$, which agrees
with the measurement of the FORTE \cite{Tierney}. The radiation intensity
of a Boltzmann-distributed bunch is about 67\% of the monoenergetic
one. We also can obtain the radiation power by $T^{-1}\int W_{\infty}(\theta)d\Omega.$
In Fig. 3c, the monoenergetic 7MeV bunch has a radiation power of 82kW. Since the
radiation power is proportional to $N^{2}$, it will increase to 1MW for $N=1.8\times10^{12}$.
These quantities of intensity and power are applicable for perfect conductors or metal. For a specific surface
(soil or sea), one should multiply them by the factor $\mathcal{R}$.

As shown in Figs 3b and 3c, the peak intensity
and power scale with $\gamma_{0}^{2}$ and $\ln\gamma_{0}$, respectively,
which is the same as transition radiation of a single electron (see Methods). With increasing electron energy, the radiation cone narrows as $\gamma_{0}^{-1}$, and the peak intensity strengthens rapidly by $\gamma_{0}^{2}$. As a result, the radiation power/energy increases slowly with $\gamma_{0}$.

The power of pairs consisting of two 100ns pulses is typically
two orders of magnitude weaker than those comprising 2-$4\mu$s
modulated pulses \cite{Jacobson3}. According to the transition-radiation model, the isolated electron bunch in the 100ns case should
contain about ten times fewer electrons than the discrete ones in the train
of electron bunches. This is in accordance with the fact that the multi-spike
x-ray bursts \cite{Howard2} are stronger than the single-spike ones. The
multi-spike x-ray bursts may origin from a higher-voltage stepped
leader, which prefers multiple breakdowns during the step formation
and also produces more relativistic electrons in each breakdown.

Our model infers that the pulse pairs could be detected on the
ground. As sketched in Fig. 1b, when the electrons are obliquely incident
on a slope, the radiation cone would center around the specular direction,
and the maximum radiation can travel along the ground surface. Actually, soon after the discovery of the TIPPs, the pulse pairs are claimed
to be observed on the ground \cite{Smith2} with the same characteristics
as the TIPPs. However, this observation is finally ascribed to an
anthropogenic source \cite{Smith1}. Therefore, ground detection
of the pulse pairs remains open and is strongly suggested.

Owing to the sporadic nature of lightning, there are only about twenty ground detections of x-rays from the stepped leaders, which limit the interpretation of electron acceleration in lightning. Coherent transition radiation has been a standard diagnostic tool for measuring three dimensional structures of electron bunches \cite{Shibata}. Extensive measurement of the pulse pairs may further shed light on the mechanism and properties of relativistic electrons generation in lightning. To avoid the signal dispersion and mode split in the ionosphere, one can observe the pairs by balloon-carried detectors in the stratosphere.

In conclusion, we reveal that the TIPPs are mainly produced by relativistic
electrons from the stepped leader of lightning, and the model can successfully
explain many satellite-observed features of the pulse pairs. The
proposed radiation mechanism is distinct from the reflection hypothesis.
Separating pulse pairs from compact intracloud discharges will
narrow the modeling of the latter. As a signal of lightning activity,
our result shows that the pulse pairs are more easily detected in
space. They can be used for real-time assessment of storms and hurricanes
across the globe by GPS satellites \cite{Hamlin}. Finally, the TIPP
source is on the ground and its potential hazards deserve attentions.

\textbf{Methods}

\textbf{Transition radiation.} Transition radiation occurs when
electrons penetrate a medium surface. For a single electron normally
incident on the surface of a perfect conductor ($\varepsilon=\infty$),
the energy distribution of transition radiation in
frequency and angle \cite{Landau} has
\begin{equation}
W_{1,\infty}(\omega,\theta)=\frac{r_{e}mc}{\pi^{2}}\frac{\beta^{2}\sin^{2}\theta}{(1-\beta^{2}\cos^{2}\theta)^{2}}.
\end{equation}
The maximum radiation is at the angle of $\theta_{\max}=\arcsin(1/\beta\gamma)\approx1/\gamma$
with $W_{1,\infty}(\omega,\theta_{\max})=\frac{r_{e}mc}{4\pi^{2}}\gamma^{2}$.
Since Eq. (3) is independent of $\omega$, the spectrum is flat. The
total radiation energy has $W_{1,\infty}\propto\intop W_{1,\infty}(\omega,\theta)d\Omega=\frac{r_{e}mc}{\pi}\left[\frac{(1+\beta^{2})}{2\beta}\ln\left(\frac{1+\beta}{1-\beta}\right)-1\right]\approx\frac{2r_{e}mc}{\pi}\ln\gamma$,
where $d\Omega=\sin\theta d\theta d\varphi$ is the differential solid
angle, and $\varphi$ is the azimuth angle. Here, all the approximations
are made for $\gamma\gg1$.

For a bunch of electrons, transition radiation can be coherent for
an electromagnetic component with wavelength longer than the bunch
size. For a Gaussian electron bunch, the energy distribution of coherent
transition radiation in frequency and angle \cite{ZhengJ} is given by
\begin{equation}
W_{\infty}(\omega,\theta)=\frac{r_{e}mc}{\pi^{2}}N^{2}e^{-\frac{\omega^{2}}{c^{2}}\sigma_{t}^{2}\sin^{2}\theta}[\intop\frac{\beta\sin\theta e^{-\frac{\omega^{2}}{2V^{2}}\sigma_{l}^{2}}}{1-\beta^{2}\cos^{2}\theta}f(p)dp]^{2},
\end{equation}
where $f(p)$ is the distribution function of the electron
momentum $p$, and fulfills $\intop_{0}^{\infty}f(p)dp=1$. For a
monoenergetic bunch with $f(p)=\delta(p-p_{0})$, Eq. (4) leads to
Eq. (1). The Boltzmann distribution $k_{0}^{-1}\exp(-k_{e}/k_{0}$$)$
has $f(p)=\frac{mc^{2}}{k{}_{0}}\frac{p}{\sqrt{1+p^{2}}}\exp(-\frac{\sqrt{1+p^{2}}-1}{k_{0}/mc^{2}}),$
where $k_{e}=(\gamma-1)mc^{2}$ is the electron kinetic energy, and
the momentum $p=P/mc=\gamma\beta$ has been normalized. We numerically calculate Eq. (4) for Boltzmann-distributed bunches.

For a general medium, the expression of radiation distribution for
a single electron is very complex \cite{Landau}. In the limit of
$\gamma\gg1$, it can be approximatively written as $W_{1,\varepsilon}(\omega,\theta)\approx\mathcal{R}(\varepsilon)W_{1,\infty}(\omega,\theta),$
where $\mathcal{R}=\left\vert (\sqrt{\varepsilon}-1)/(\sqrt{\varepsilon}+1)\right\vert ^{2}$
is the Fresnel reflectivity. Accordingly, one has $W_{\varepsilon}(\omega,\theta)\approx\mathcal{R}(\varepsilon)W_{\infty}(\omega,\theta)$
for the coherent transition radiation. The appearance of $\mathcal{R}$
is because self-fields of a relativistic bunch are predominantly
transverse and coherent transition radiation can be understood as
the reflection of the bunch field by the surface \cite{Wu}.

\begin{acknowledgements} This work was supported by the Thousand
Youth Talents Plan, NSFC (No. 11374262), and Fundamental Research
Funds for the Central Universities. \end{acknowledgements}

\end{document}